\documentclass{ws-procs975x65}

\def\beq{\begin{equation}}
\def\eeq{\end{equation}}

\begin{document}

\title{Gravitational waves in $\alpha-$attractors}
\author{K. Sravan Kumar$^*$
, Jo\~ao Marto $^\dagger$ 
and Paulo Vargas Moniz$^\ddagger$
}

\address{Departamento de F\'{i}sica, Universidade da Beira Interior\\
Covilh\~a, 6200, Portugal\\
Centro de Matem\'atica e Aplica\c{c}\~oes da Universidade da Beira Interior
(CMA-UBI)\\
 $^*$E-mail: sravan@ubi.pt\\
 $^\dagger$E-mail: jmarto@ubi.pt\\
 $^\ddagger$E-mail: pmoniz@ubi.pt
}

\author{Suratna Das} 

\address{Indian Institute of Technology\\
Kanpur, 208016, India \\
 E-mail: suratna@iitk.ac.in
}

\begin{abstract}
We study inflation in the $\alpha-$attractor model under a non-slow-roll
dynamics with an ansatz proposed by Gong \& Sasaki \cite{Gong:2015ypa}
of assuming $N=N\left(\phi\right)$. Under this approach, we construct
a class of local shapes of inflaton potential that are different from
the T-models. We find this type of inflationary scenario predicts
an attractor at $n_{s}\sim0.967$ and $r\sim0.00055$. In our approach,
the non-slow-roll inflaton dynamics are related to the $\alpha-$parameter
which is the curvature of K\"ahler geometry in the SUGRA embedding of
this model. 
\end{abstract}

\keywords{inflation, supergravity , $ \alpha- $attractors.}

\bodymatter

\section{Introduction}

Inflationary cosmology has become an extremely convincing theory of
the early universe concerning the recent release of Planck data \cite{Ade:2015lrj}.
We now have stringent bounds on spectral index $n_{s}=0.968\pm0.006$
at 95\% CL and also for the tensor scalar ratio, which is severely
bounded $r<0.11$. Among the broad variety of inflationary scenarios,
the Starobinsky model, with $R+R^{2}$ term, and Higgs inflation \cite{Starobinsky:1980te,Bezrukov:2007ep}
stands in a privileged region in the middle of $\left(n_{s},r\right)$
plane \cite{Ade:2015lrj}. Moreover, the Starobinsky model prediction
is identified as a target spot for the predictions in many inflationary models,
i.e.,

\begin{equation}
n_{s}=1-\frac{2}{N}\quad r=\frac{12}{N^{2}}\:.\label{sweetspot}
\end{equation}
Since the first release of Planck 2013, these two models (Starobinsky
and Higgs) started to attract a lot of attention and became extensively
studied and realized in the context of conformal symmetries and later
generalized as $\alpha-$ and non-minimal (or) $\xi-$ attractors,
in addition these models have been embedded in supergravity (SUGRA).
These two classes of models have also, posteriori, been unified as
cosmological attractor models (CAM) \cite{Galante:2014ifa,Cecotti:2014ipa,Roest:2013fha}.

In this work we present a non-slow-roll inflation dynamics in the
$\alpha-$attractor model using the recently proposed approach of
Gong and Sasaki (GS) \cite{Gong:2015ypa}. More concretely, we mainly
focus on non-canonical aspect of $\alpha-$ attractor model and explore
different class of inflaton potentials under non-slow-roll dynamics. We show that considering a non-slow-roll dynamics, the $\alpha-$attractor
model remains compatible with any value of $r<0.1$. And our study predicts an attractor at $n_{s}\approx0.967$ and $r\approx5.5\times10^{-4}$ which are very close to the predictions of the first chaotic inflationary model in supergravity (Goncharov-Linde model) \cite{Goncharov:1983mw}.
 
\section{$\alpha-$attractor model}

\label{alphaattractor}The Lagrangian for $\alpha-$attractor models in the Einstein frame
\cite{Galante:2014ifa,Kallosh:2013yoa} is given by

\begin{equation}
\mathcal{\mathcal{L}}_{E}=\sqrt{-g}\left[\frac{R}{2}-\frac{1}{\left(1-\phi^{2}/6\alpha\right)^{2}}\frac{\left(\partial\phi\right)^{2}}{2}-f^{2}\left(\phi/\sqrt{6\alpha}\right)\right]\:.,\label{alphaL}
\end{equation}
This model is first realized in the context of two field models with
spontaneously broken conformal invariance. In order to prevent the
negative gravity in Jordan frame, it is requred to satisfy $\vert\phi\vert<\sqrt{6\alpha}$.
Furthermore, in the SUGRA embedding of this model, the parameter $\alpha$
is shown to be related to the curvature of K\"ahler manifold as
\begin{equation}
\mathcal{R}_{\mathcal{K}}=-\frac{2}{3\alpha}\label{kalhercurvature}
\end{equation}
The scalar field Lagrangian in Eq.(\ref{alphaL}) is a subclass of k-inflationary
model where the kinetic term is linear%
\footnote{$K\left(\phi\right)=1$ gives the canonical kinetic term.%
} in $X$, i.e.,

\begin{equation}
P\left(X,\phi\right)=K\left(\phi\right)X-f^{2}\left(\phi/\sqrt{6\alpha}\right)\,,\label{kinflationE}
\end{equation}
where $K\left(\phi\right)=\frac{1}{\left(1-\phi^{2}/6\alpha\right)^{2}}$
and $X=-\frac{\left(\partial\phi\right)^{2}}{2}$. The speed of sound
for these class of models is $c_{s}^{2}=1$, therefore
these models are not expected to show large non-Gaussianities.

In this theory, the Friedmann equation is

\begin{equation}
H^{2}=\frac{1}{3}\left(XK\left(\phi\right)+f^{2}\left(\frac{\phi}{\sqrt{6\alpha}}\right)\right)\,.\label{Efriedmann}
\end{equation}
The Raychaudhuri equation is

\begin{equation}
\dot{H}=-XP_{,X}\,\:\: \text{with}\:\: P_{,X}= \frac{\partial P}{\partial X},\label{RaychaudhuriE}
\end{equation}
and the equation of motion for the scalar field is given by

\begin{equation}
\frac{d}{dt}\left(K\left(\phi\right)\dot{\phi}\right)+3HK\left(\phi\right)\dot{\phi}-P_{,\phi}=0\,.\label{Eom}
\end{equation}
In the literature it is found that inflation in the $\alpha-$attractor
model has been realized in terms of a canonically normalized field $\left(\varphi\right)$ as

\begin{equation}
\frac{d\varphi}{d\phi}=\frac{1}{\left(1-\frac{\phi^{2}}{6\alpha}\right)}\Rightarrow\frac{\phi}{\sqrt{6\alpha}}=\tanh\frac{\varphi}{\sqrt{6\alpha}}\,.\label{canonical field}
\end{equation}
The slow-roll inflationary predictions
of $\alpha-$attractor models are

\begin{equation}
n_{s}=1-\frac{2}{N}\quad r=\frac{12\alpha}{N^{2}}\,.\label{sweetspot-1}
\end{equation}
In terms of this canonically normalized field $\left(\varphi\right)$ the
equation of motion (\ref{Eom}) becomes

\begin{equation}
\ddot{\varphi}+3H\dot{\varphi}+V_{,\varphi}=0\label{canonicalEOM}
\end{equation}
Therefore, under slow-roll assumption this reduces to

\begin{equation}
3H\dot{\varphi}\simeq V_{,\varphi}\label{slowroll}
\end{equation}

Our interest relies on the possibility of obtaining viable inflationary
predictions within this model with non-slow-roll dynamics. Therefore, in the present work, we restrict
ourselves to the study of the original scalar field dynamics  within
the allowed range $\phi^{2}<6\alpha$.

\subsection{Non-slow-roll dynamics }
\label{non-slow-roll-dynamics}The recent work by Gong \& Sasaki (GS)
\cite{Gong:2015ypa} points out a cautionary remark on applying slow-roll
in the context of k-inflation. The argument, presented there, lies
in the fact that the second derivative term in the equation of motion
(Eq. \ref{Eom}) may not be negligible in general. They have proposed an ansatz for scalar field and illustrated non-slow-roll inflation in the context of non-canonical models.

In the $\alpha-$attractor model, since the original scalar field $\phi$ is non-canonical
(see Eq. (\ref{kinflationE})), we assume the following ansatz during inflation%
\footnote{We start with a similar parametrization as the one used in section
3.2 of \cite{Gong:2015ypa}.%
}

\begin{equation}
\phi=n\exp\left(\beta N\right)\,,\label{sasakiparametrization}
\end{equation}
where $N=\ln a\left(t\right)$ is the number of efoldings counted
backward in time from the end of inflation and $n$ is treated as
a free parameter that specifies the value of the field at $N\rightarrow0$.
We assign Eq. (\ref{sasakiparametrization}) as GS parametrization
for subsequent reference.   

Substituting $\phi$ from Eq. (\ref{sasakiparametrization})
in the Raychaudhuri equation we obtain

\begin{equation}
H^{\prime}=\frac{\alpha^{2}H\left(N\right)}{2}\phi^{2}K\left(\phi\right)\,,\label{rayintegrate}
\end{equation}
where the prime  $ ^{\prime} $ denotes differentiation with respect to $N$. Integrating
Eq. (\ref{rayintegrate}), we get

\begin{equation}
H\left(N\right)=\lambda e^{-\frac{9\beta\alpha^{2}}{\phi^{2}-6\alpha}}\,,\label{HEsol}
\end{equation}
where $\lambda$ is the integration constant. At this point, we should mention that our calculations are similar to the Hamilton-Jacobi like formalism.

Inserting the aforementioned
solution in the Friedmann Eq. (\ref{Efriedmann}), we can express the
local shape of the potential during inflation as

\begin{equation}
f^{2}\left(\frac{\phi}{\sqrt{6\alpha}}\right)=\lambda e^{-\frac{9\beta\alpha^{2}}{\phi^{2}-6\alpha}}\left(3-\frac{\beta^{2}\phi^{2}}{2\left(1-\frac{\phi^{2}}{6\alpha}\right)^{2}}\right)\,.\label{potential}
\end{equation}

The suitable choice of potentials considered in the case of slow-roll $ \alpha- $attractors are power law type $ V\sim\phi^{2n} $ in terms of original scalar field (or) T-models, i.e.,$  \,V\sim\tanh^{2n}{\frac{\varphi}{\sqrt{6\alpha}}} $ in terms of canonically normalized field \cite{Galante:2014ifa,Kallosh:2013yoa}. We can deduce from Eq. (\ref{potential}) that the local shape of the potential in non-slow-roll approach is different from the power-law (or) T-models. Our study about the non-slow-roll approach widens the scope for different shapes of inflationary potentials in $ \alpha- $ attractors.  

Subsequently, we write the slow-roll parameters general definitions as%

\begin{equation}
\epsilon=\frac{H^{\prime}}{H}\quad,\quad\eta=-\frac{\epsilon^{\prime}}{\epsilon}\,.\label{epsilon}
\end{equation}
Substituting Hubble parameter from Eq. (\ref{HEsol}) in the slow-roll
parameter and demanding the end of inflation $\epsilon=1$ at $N=0$
we get

\begin{equation}
\alpha=\frac{n^{2}}{3\sqrt{2}\beta n+6}\,.\label{alphafix}
\end{equation}
Consequently constraining the parameter space $\left(n,\,\beta\right)$
automatically gives the values of $\alpha$. From Eqs. (\ref{sasakiparametrization}),
(\ref{potential}) and (\ref{alphafix}), we can say that the local
shape of the potential, the inflaton dynamics and the parameter $\alpha$
are interconnected. In other words, identifying $\alpha$ as the curvature
of K\"ahler geometry given by Eq. (\ref{kalhercurvature}), we can establish
a web of relations 

\vspace{0.5cm}
\hspace{0.05cm}
\begin{picture}(0,0)%
   
    \put(65,0){\color[rgb]{0,0,0}\makebox(0,0)[lb]{K\"ahler Geometry}}%
    \put(165,0){\makebox(0,0)[lb]{{\Huge{$\leftrightharpoons$}}}}%
    \put(205,0){\color[rgb]{0,0,0}\makebox(0,0)[lb]{Inflaton Dynamics}}%
    
    \put(115,-15){\color[rgb]{0,0,0}\rotatebox{-45}{\makebox(0,0)[lb]
    {\Huge{$\leftrightharpoons$}}}}%
    \put(225,-5){\color[rgb]{0,0,0}\rotatebox{-135}{\makebox(0,0)[lb]
    {\Huge{$\leftrightharpoons$}}}}%
     \put(125,-45){\color[rgb]{0,0,0}\makebox(0,0)[lb]{Local shape of the potential}}%

\end{picture}%

\vspace{2cm}

From the above schematic diagram we can decipher that the class of potentials which are obtained by allowing different values for
$\left(n,\,\beta\right)$ is related to the family of K\"ahler geometries,
which determine the dynamics of inflaton during inflation. 

\section{Non-slow-roll $\alpha-$ attractors}

In this section, we present the inflationary predictions of $ \alpha- $  attractor model in the context of non-slow-roll \cite{Kumar:2015mfa}. 

The power spectrum of primordial curvature perturbation is

\begin{equation}
{\cal P}_{\zeta}=\frac{\gamma_{s}}{2}\frac{H^{2}}{4\pi^{2}\epsilon}\biggr\lvert_{k=aH}\quad,\quad\gamma_{s}\equiv2^{2\nu_{s}-3}\frac{\Gamma\left(\nu_{s}\right)^{2}}{\Gamma(3/2)^{2}}\left(1-\epsilon\right)^{2}\,.\label{pwrspectrum}
\end{equation} 
From the Planck data \cite{Ade:2015lrj} the power spectrum amplitude
is known to be ${\cal P}_{\zeta_{*}}=2.2\times10^{-9}$. Using this
bound, with Eqs. (\ref{HEsol}) and (\ref{pwrspectrum}), we constrain
$\lambda\sim\mathcal{O}\left(10^{-6}\right)$.

The scalar spectral index is given by

\begin{equation}
n_{s}-1=3-2\nu_{s}\,.\label{ns-1}
\end{equation}
where 
\begin{equation}
\begin{aligned}\nu_{s}= & \left(\frac{3}{2}+\epsilon+\epsilon^{2}+\epsilon^{3}\right)+\left(\frac{1}{2}+2\epsilon+\frac{29\epsilon^{2}}{6}+\frac{82\epsilon^{3}}{9}\right)\eta\\
 &+\left(-\frac{1}{6}+\frac{23\epsilon}{18}+\frac{1069\epsilon^{2}}{108}+\frac{5807\epsilon^{3}}{162}\right)\eta^{2}+ +\mathcal{O}\left(\epsilon^{3}\,,\,\eta^{3}\right)\,.
\end{aligned}
\label{nusfull}
\end{equation}
The power spectrum of primordial tensor perturbation is 

\begin{equation}
{\cal P}_{T}=8\gamma_{t}\frac{H^{2}}{4\pi^{2}}\biggr\lvert_{k=aH}\quad,\quad\gamma_{t}\equiv2^{2\nu_{t}-3}\frac{\Gamma\left(\nu_{t}\right)^{2}}{\Gamma(3/2)^{2}}\left(1-\epsilon\right)^{2}\,.\label{ptspectrum}
\end{equation}
The tensor tilt $\left(n_{t}\right)$ is given by 
\begin{equation}
n_{t}=3-2\nu_{t}\,.\label{nt}
\end{equation}
Calculating $\nu_{t}$ up to the second order in the slow-roll parameters
\begin{equation}
\begin{aligned}\nu_{t}= & \left(\frac{3}{2}+\epsilon+\epsilon^{2}+\epsilon^{3}\right)+\left(\frac{4\epsilon}{3}+\frac{37\epsilon^{2}}{9}+\frac{226\epsilon^{3}}{27}\right)\eta+\\
 & +\left(\epsilon+\frac{227\epsilon^{2}}{27}+\frac{875\epsilon^{3}}{27}\right)\eta^{2}+\mathcal{O}\left(\epsilon^{3}\,,\,\eta^{3}\right)\,.
\end{aligned}
\label{nut}
\end{equation}
From Eqs.~(\ref{pwrspectrum}) and (\ref{ptspectrum}) we define the tensor to scalar ratio as

\begin{equation}
r=\dfrac{\cal P_{T}}{\cal P_{\zeta}}=16\dfrac{\gamma_{t}}{\gamma_{s}}\epsilon\,.\label{TSR}
\end{equation}

\begin{figure}[ht]
\centering\includegraphics[height=1.6in]{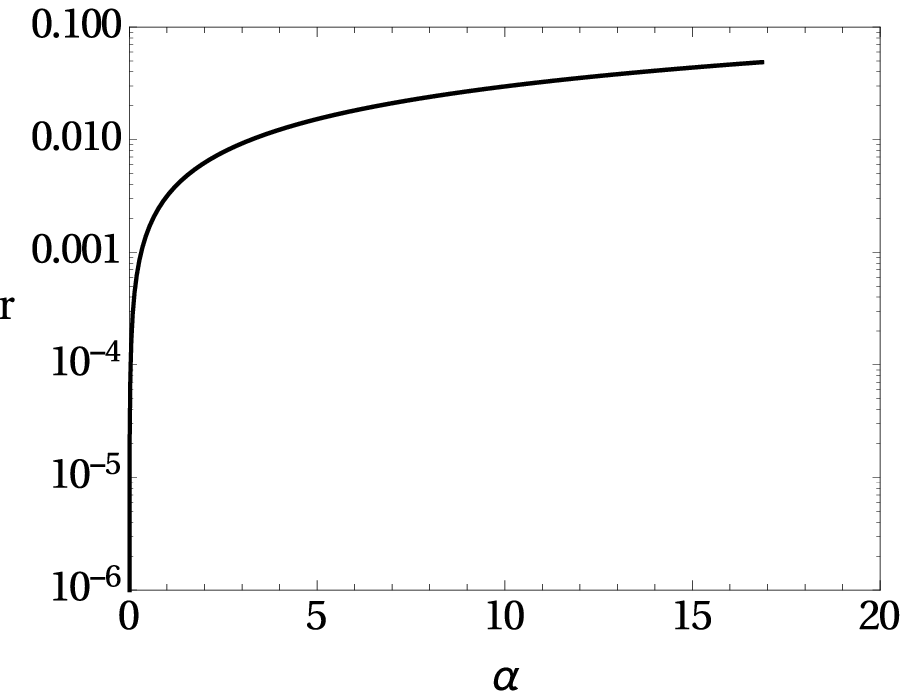}\quad{}\includegraphics[height=1.6in]{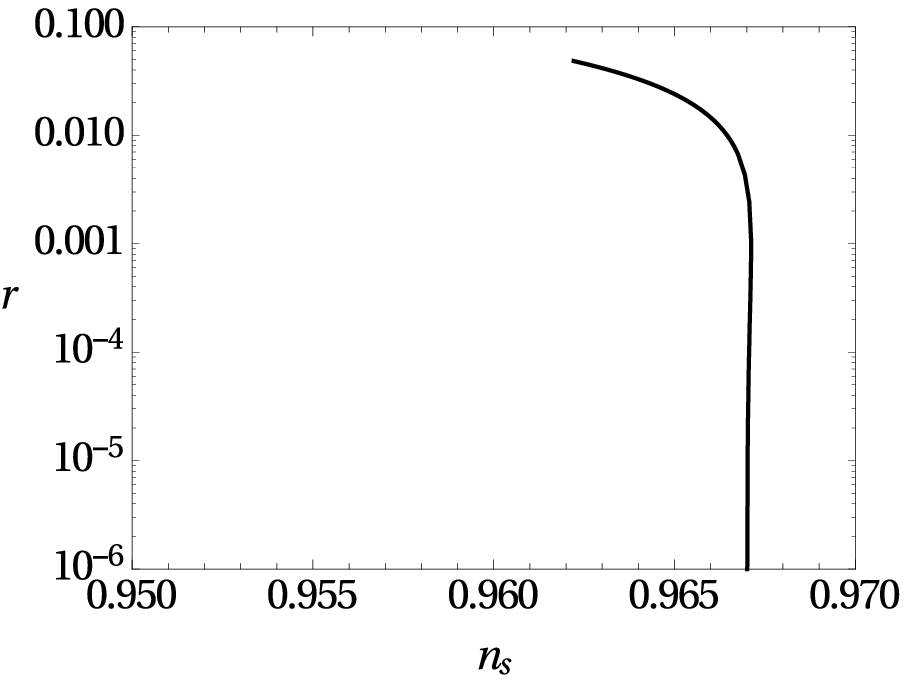}\quad{}\caption{In the left panel, we plot tensor scalar ratio $\left(r\right)$ vs $\alpha-$ parameter. In the right panel we depict parametric plot of spectral index $ \left(n_{s} \right) $ vs tensor scalar ratio $ \left(r\right) $. For both the plots we
have taken $\beta\sim-0.002$ , $0<n<10$ and $ N=60 $.}
\label{Ipredic} 
\end{figure}

In Fig. \ref{Ipredic} we present the inflationary predictions of non-slow-roll $ \alpha- $attractors by constraining the parameter space $ \left(n\,,\,\beta\right) $.  The values $ n=1 $, $ \beta=-0.002 $ leads to an attractor point $ n_{s}=0.967 $ and $ r=0.00055 $. We obtain the behaviour, $ r\to0 $ as $ n\to0 $ (or equivalently $ \alpha\to0 $) without altering the prediction of scalar tilt.

\section{Concluding remarks }

In this paper, we studied the non-slow-roll dynamics of inflation
field in $ \alpha- $attractors. We find that the the predictions of $\left(n_{s},\, r\right)$ in this case are well compatible with Planck 2015 \cite{Ade:2015lrj}. We argue that, in our case, the shape of potential during inflation is different from T-models and is naturally related to the curvature of K\"alher geometry in the SUGRA embedding of this model. 

\section{Acknowledgements}

SK acknowledges for the support of grant SFRH/BD/51980/2012 from FCT, Portugal. This research work is supported by the grants PTDC/FIS/111032/2009 and UID/MAT/00212/2013. SD acknowledges the grant  IFA13-PH-77 from DST, India. 

\bibliographystyle{h-physrev}
\bibliography{References}

\end{document}